\documentclass[pre,aps,twocolumn,amsmath,amssymb]{revtex4-1}

\pdfoutput=1
\usepackage{hyperref}
\usepackage{graphicx}
\usepackage[usenames]{color}
\usepackage{bm}
\usepackage[final]{movie15}

\definecolor{blue}{rgb}{0,0,0}
\newcommand{\bb}[1]{\textcolor{blue}{#1}}

\def\cal#1{\mathcal{#1}}
\def\eqq#1{Eq.~(\ref{#1})}
\def\eq#1{(\ref{#1})}
\def\av#1{\langle #1 \rangle}

\def\f#1{Fig.~\ref{#1}}

\def\c#1{~\cite{#1}}
\def\cc#1{Ref.\c{#1}}

\def\ta0{\tilde{a}_0}
\def\tp{\tilde{p}}

\def\s#1{Section~\ref{#1}}

\def\e{{\rm e}}

\def\beq{\begin{equation}}
\def\eeq{\end{equation}}
\def\bea{\begin{eqnarray}}
\def\eea{\end{eqnarray}}

\begin{document}

\title{Evolutionary reinforcement learning of dynamical large deviations}
\author{Stephen Whitelam$^1$}
\email{{\tt swhitelam@lbl.gov}}
\author{Daniel Jacobson$^2$}
\author{Isaac Tamblyn$^{3,4}$}
\affiliation{$^1$Molecular Foundry, Lawrence Berkeley National Laboratory, 1 Cyclotron Road, Berkeley, CA 94720, USA\\
$^2$Division of Chemistry and Chemical Engineering, California Institute of Technology, Pasadena, California 91125, USA\\
$^3$University of Ottawa \& $^4$National Research Council of Canada, Ottawa, ON, Canada}

\begin{abstract}
We show how to calculate the likelihood of dynamical large deviations using evolutionary reinforcement learning. An agent, a stochastic model, propagates a continuous-time Monte Carlo trajectory and receives a reward conditioned upon the values of certain path-extensive quantities. Evolution produces progressively fitter agents, eventually allowing the calculation of a piece of a large-deviation rate function for a particular model and path-extensive quantity. For models with small state spaces the evolutionary process acts directly on rates, and for models with large state spaces the process acts on the weights of a neural network that parameterizes the model's rates. This approach shows how path-extensive physics problems can be considered within a framework widely used in machine learning.
\end{abstract}

\maketitle

\section{Introduction} 

Machine learning provides the physics community with methods that complement the traditional ones of physical insight and manipulation of equations. Many-parameter ans\"atze, sometimes encoded in the form of neural networks, can learn connections between physical properties (such as the positions of atoms and a system's internal energy) without drawing upon an underlying physical model\c{behler2007network,mills2017deep,ferguson2018machine,artrith2018constructing,singraber2018density,desgranges2018new,thurston2018machine,singraber2019library,han2016deep,schutt2017dqtn,yao2018tensormol,schutt2018schnet,carrasquilla2017phases,portman2017sampling,rem2019experiment}. Reinforcement learning is a branch of machine learning concerned with performing actions so as to maximize a numerical reward\c{sutton2018reinforcement}. It has a close connection to ideas of stochastic control enacted by variational or adaptive algorithms\c{ahamed2006adaptive,basu2008learning,borkar2010learning,borkar2003peformance,chetrite2015variational,kappen2016adaptive,nemoto2017finite,ferre2018adaptive,PhysRevE.98.063303}. A recent success of reinforcement learning is the playing of computer games\c{QL,DQN,Atari,Atari2600,Actor,DeepMind,MuJoCo,MDP,Rogue,NFQ,Soccer,PPO,Guber,OpenAI,VizDoom,VizDoom2,Go,Go2}. Here we show that reinforcement learning can also be used to propagate trajectories of a stochastic dynamics conditioned upon potentially rare values of a path-extensive observable. Doing so allows the calculation of dynamical large deviations, which are of fundamental importance, being to dynamical quantities what free energies are to static ones\c{touchette2009large,garrahan2009first,den2008large,ellis2007entropy}.

Calculating large deviations is a challenging problem for which specialized numerical methods are required\c{giardina2006direct,ahamed2006adaptive,basu2008learning,borkar2010learning,borkar2003peformance,touchette2009large,garrahan2009first,chetrite2015variational,ray2018exact,nemoto2017finite,ferre2018adaptive,banuls2019using}. Here we work within the framework of the VARD (variational ansatz for rare dynamics) method of \cc{jacobson2019direct}. VARD focuses on the ratio of probabilities with which a given dynamical trajectory can be generated by two stochastic models, the second model being an ansatz for the behavior of the first model when conditioned upon a particular value of a time-extensive observable. We showed previously that simple, physically motivated choices for the second model can be used to bound the likelihood of rare events in the first, and, if certain criteria are fulfilled,  calculate this likelihood exactly\c{jacobson2019direct}. That paper contains details of the convergence criteria and statistical errors associated with the VARD method. In the present paper we use evolutionary reinforcement learning to calculate bounds associated with multi-parameter ans\"atze, in some cases encoded by neural networks, and show that these bounds are tighter than bounds associated with the few-parameter, physically motivated ans\"atze used in \cc{jacobson2019direct}. Moreover, by direct comparison with answers obtained by other means we show that bounds derived from evolutionary learning already provide a very good approximation of the log-likelihood of rare events in systems for which state-of-the-art methods must be used, showing the present approach to have the potential to address cutting-edge problems. The evolutionary reinforcement learning procedure we describe is conceptually and technically simple, and does not require insight into the models under study or access to the formal results of large-deviation theory. It therefore offers an alternative to existing approaches, and provides an example of one of the potentially large number of applications of reinforcement learning in physics. 

\section{Large deviations by change of dynamics}

 To set the large-deviations problem in a form amenable to reinforcement learning, consider a continuous-time Monte Carlo dynamics on a set of discrete states, with $W_{xy}$ the rate for passing between states $x$ and $y$, and $R_x = \sum_{y \neq x} W_{xy}$ the escape rate from $x$\c{gillespie1977exact}. This dynamics generates a trajectory $\omega=x_0 \to x_1 \to \dots \to x_{N(\omega)}$ consisting of $N(\omega)$ jumps $x_n \to x_{n+1}$ and associated jump times $\Delta t_n$. In the language of reinforcement learning, $W_{xy}$ is a policy (often denoted $\pi$) that stochastically selects a new state and a jump time given a current state.

Stochastic trajectories can be characterized by path-extensive observables $A=aT$, with
 \beq
 \label{adef}
 a = T^{-1} \sum_{n=0}^{N-1} \alpha_{x_n x_{n+1}}.
 \eeq
Here $\alpha_{xy}$ is the change of the observable upon moving between $x$ and $y$. This type of observable describes many physically important quantities, including work, entropy production, and non-decreasing counting observables\c{seifert2005entropy,speck2012large,lecomte2010current,garrahan2009first,fodor2015activity}. Let the typical value of $a$ be $a_0$, the limiting value of \eq{adef} for a long trajectory of the model $W_{xy}$. Finite-time fluctuations $a \neq a_0$ occur with a probability controlled by the distribution $\rho_T(A)$, taken over all trajectories of length $T$. For large $T$ this distribution often adopts the large-deviation form\c{den2008large,touchette2009large}
\beq
\label{d1-main}
\rho_T(A) \approx \e^{-T J(a)}.
\eeq 
$J(a)$ is the large-deviation rate function, which quantifies the likelihood of observing atypical values of $a$\c{touchette2009large,den2008large}. Calculation of $J(a)$ far from $a_0$ using only the original model is not feasible, because such values of $a$ occur rarely. Instead, we can consider a new stochastic model, which we call the reference model, whose purpose is to allow the calculation of $J(a)$ potentially far from $a_0$\c{bucklew2013introduction,chetrite2015variational}. 

With the reference model we can carry out a form of importance sampling\c{ahamed2006adaptive,basu2008learning,borkar2010learning,borkar2003peformance,den2008large,glynn1989importance,sadowsky1990large,bucklew1990monte,bucklew1990large,asmussen2007stochastic,juneja2006rare,bucklew2013introduction,touchette2009large,jacobson2019direct}. Let the rates of the reference model be $\tilde{W}_{xy}$ and $\tilde{R}_x = \sum_{y \neq x} \tilde{W}_{xy}$, and let the limiting value of \eq{adef} for a long reference-model trajectory be $\ta0$. Then an upper bound on $J(a)$ at $a=\ta0$ is given by the value of
\beq
\label{j0}
J_0= -T^{-1} \sum_{n=0}^{N-1} q_{x_n x_{n+1}}
\eeq
for a long reference-model trajectory, where
\beq
\label{j0b}
q_{x_n x_{n+1}} = \ln \frac{W_{x_n x_{n+1}}}{\tilde{W}_{x_n x_{n+1}}}-\tilde{\Delta t}_n (R_{x_n} - \tilde{R}_{x_n}).
\eeq
Here $\tilde{\Delta t}_n = -\ln \eta/\tilde{R}_{x_n}$ is the jump time of the reference model, and $\eta$ is a random number uniformly distributed on $(0,1]$. \eqq{j0} follows from straightforward algebra (see Appendix A). It can be motivated by noting that the probability of a jump $x \to y$ in time $\tilde{\Delta t}$ occurs in the reference model with probability $\tilde{W}_{xy} \e^{-\tilde{R}_x \tilde{\Delta t}}$, and in the original model with probability $W_{xy} \e^{-R_x \tilde{\Delta t}}$; \eqq{j0} is the sum over a trajectory of the log-ratio of such terms.

Our aim in this paper to use evolutionary reinforcement learning to find a reference model (a new policy) $\tilde{W}_{xy}$ that produces a particular typical value of \eq{adef}, say $\tilde{a}_0$, and which minimizes \eq{j0}. 

Given a value of $\tilde{a}_0$, the model associated with the smallest possible value of \eq{j0} is called the driven or effective model, and its typical behavior is equivalent to the conditioned rare behavior of the original model\c{chetrite2015variational}. The typical behavior of the driven model yields, from \eq{j0}, the piece $J(\tilde{a}_0)$ of the rate function of the {\em original} model at the point $a=\tilde{a}_0$. In previous work\c{jacobson2019direct} we showed that the reference model need only be {\em close} to the driven model (in a sense made precise in that paper) in order to calculate $J(\tilde{a}_0)$; from the typical behavior of such a reference model we get a bound [\eqq{j0}] $J_0(\tilde{a}_0)>J(\tilde{a}_0)$, and by sampling the atypical behavior of the reference model we can (under certain conditions) compute a correction $J_1(\tilde{a}_0)$ such that $J_0(\tilde{a}_0)+J_1(\tilde{a}_0)=J(\tilde{a}_0)$. There we showed that simple, physically-motivated choices of reference model lead to relatively tight bounds ($J_0 \approx J$) and small corrections $J_1 \ll J_0$ for a set of models taken from the literature. In this paper we show how to further improve the quality of these bounds using multiparameter ans\"atze determined by evolutionary learning. We do not address here the calculation of the correction term (for more detail of that calculation, including convergence criteria, see \cc{jacobson2019direct}), but for the cases studied here the correction term is so small that the bound alone suffices for the purposes of plotting the rate function. Error bars associated with the bound scale as $1/\sqrt{N}$, where $N$ is the number of events in the trajectory. For the models studied here we chose $N$ large enough that error bars are smaller than symbol sizes.

The formulation of this section describes an extreme example of reinforcement learning in which there is no instantaneous reward, only an overall reward (or return) associated with the entire trajectory\c{sutton2018reinforcement}. Given that we possess a constraint on $a$ and work in continuous time, this problem also falls outside the (standard) Markov decision process framework\c{li2013basic,singh2007policy}. A natural approach to such problems are evolutionary algorithms, which are simple to apply and have been shown to be competitive with gradient-based methods\c{sutton2018reinforcement} on complex problems whose solution requires upwards of thousands of parameters\c{GA,GA2,lehman2018more,salimans2017evolution,zhang2017relationship,lehman2018safe,conti2018improving,Guber}. We use an evolutionary approach in this paper.

\section{Large deviations via evolutionary reinforcement learning}
\label{fourstate}
\begin{figure}[] 
   \centering
  \includegraphics[width=\linewidth]{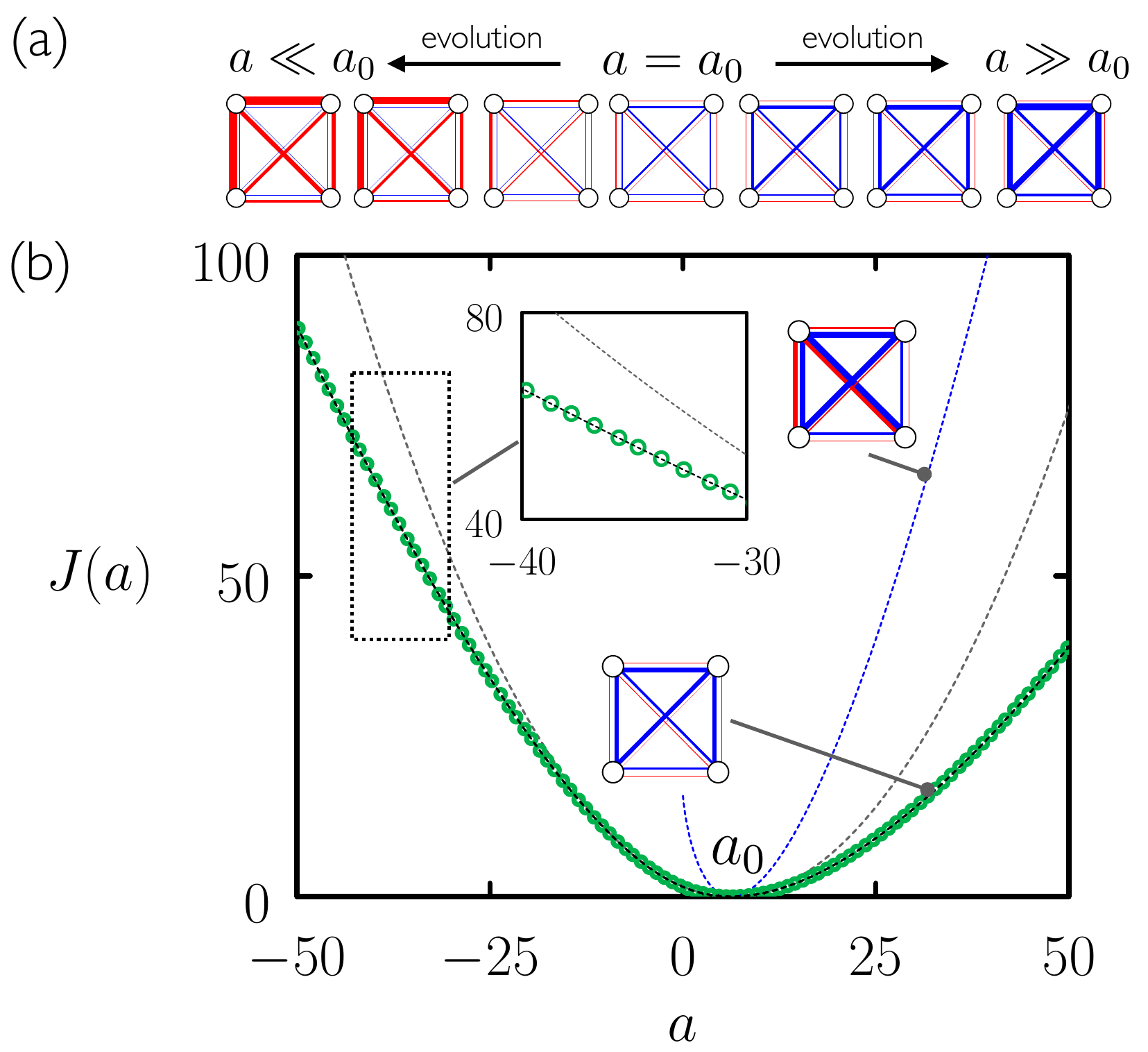} 
   \caption{(a) Evolutionary reinforcement learning can produce versions of the 4-state model of \cc{gingrich2016dissipation} whose typical dynamics are exactly equivalent to the rare dynamics of the original model (center) conditioned on values of entropy production $a$. (b) From these we can calculate the corresponding large-deviation rate function, $J(a)$. The black dashed line is the exact answer, obtained by matrix diagonalization\c{den2008large,touchette2009large}, and the blue- and gray dashed lines are the Conway-Maxwell-Poisson bound\c{garrahan2017simple} and the universal current bound\c{pietzonka2016universal,gingrich2016dissipation}, respectively. The green points describe a bound resulting from a set of models generated by evolutionary reinforcement learning; this bound is effectively exact. Each green point is calculated using a single trajectory of a stochastic model produced by the evolutionary process. Inset left: enlargement of the boxed area. Inset right: we contrast one model (lower image) produced by evolution (see panel (a)) with a second model (upper image) that produces the same typical value of $a$ but whose rates are uniformly scaled versions of the original model.}
   \label{fig1}
\end{figure}

As proof of principle we consider the example of entropy production in the 4-state model of \cc{gingrich2016dissipation}. The model's rates do not satisfy detailed balance, and so it produces nonzero entropy on average~\footnote{The model's rates are $W_{12} = 3$, $W_{13} = 10$, $ W_{14} = 9$, $ W_{21} =10$, $ W_{23} = 1$, $ W_{24}  = 2$, $ W_{31} = 6$, $ W_{32} = 4$, $ W_{34} =1$, $ W_{41} = 7$, $ W_{42} = 9$, and $ W_{43} = 5$.}. The dynamical observable $a$ is \eq{adef} with $\alpha_{xy} = \ln (p_{xy}/p_{yx})$, where $p_{xy} = W_{xy}/R_x$. In \f{fig1}(a) we depict the model (the middle picture), with states $x$ numbered clockwise from 1 at the top left. Red and blue links denote connections $x\to y$ with negative and positive entropy production, respectively, and the thickness of the links is proportional to the rate associated with the connection. The model's state space is small enough that the master operator can be solved by diagonalization\c{touchette2009large}, yielding the exact rate function $J(a)$, shown as a black dashed line in \f{fig1}(b).

We can reconstruct this function using evolutionary reinforcement learning by mutating the rates $\tilde{W}_{xy}$ of a set of reference models until desired values of \eq{adef} and \eq{j0} are achieved. The process is as follows.

We start by running a trajectory of the reference model (of $N=10^4$ events) and recording the typical value of the observable and bound, the long-time limits of \eq{adef} and \eq{j0}, respectively. Initially the reference model is the original model, $\tilde{W}_{xy}=W_{xy}$, and so $a=a_0$ and $J_0=0$.

To perform an evolutionary step we create a mutant model whose rates are
\beq
\hat{W}_{xy} = \e^{\epsilon(\eta_{xy}-1/2)} \tilde{W}_{xy}.
\eeq
Here $\epsilon$ is an evolutionary rate and $\eta_{xy}$ is a uniformly distributed random number on $(0,1]$. \bb{The parameter $\epsilon$ is a learning rate and its effect is similar to other types of learning rate in machine learning, or basic step size in Monte Carlo simulation: if it is too small then we do not explore parameter space rapidly enough; if it is too large then the acceptance rate is too low; and somewhere in between these extremes its precise numerical value does not matter. The latter regime must be determined empirically, and we found values of $\epsilon$ of order 0.1 to be acceptable.} 

With this new set of rates we run a new trajectory and compute the new values of $a$ and $J_0$, called $\hat{a}$ and $\hat{J}_0$, respectively. If our selection criteria are fulfilled (see below) then we accept the mutation, and set $\tilde{W}_{xy} = \hat{W}_{xy}$, $a=\hat{a}$, and $J_0 = \hat{J}_0$ (i.e. the mutant model becomes the new reference model); if not, we retain the current reference model.

We imposed two types of selection criteria. For the first, called $a$-evolution, we accepted the mutation if $\hat{a}$ is closer than $a$ to a specified target value $a^\star$, i.e. if
\beq
\label{simple}
|\hat{a}-a^\star|<|a-a^\star|.
\eeq
 For the second, called $J$-evolution, we accept the mutation if $\hat{J}_0$ is smaller than $J_0$ {\em and} if $\hat{a}$ lies within a tolerance $\delta$ of a specified pinning value $a^\dagger$, i.e. if
 \beq
 \label{jev}
 \hat{J}_0<J_0 \quad {\rm and} \quad |\hat{a}-a^\dagger|<\delta.
 \eeq
The process of $a$-evolution leads to reference models able to generate values of $a$ far from $a_0$, while $J$-evolution leads to reference models that generate values of $a$ in a manner as close as possible to the original model. \bb{The role of the parameter $\delta$ in \eq{jev} is to constrain the reference model to a particular window of $a$, and its value can be chosen for convenience (e.g. to ensure that we plot particular points along the rate function).}

We alternated $5$ steps of $a$-evolution, using an evolutionary rate of $\epsilon=0.1$, with $50$ steps of $J$-evolution, using an evolutionary rate of $\epsilon=0.05$ and a tolerance of $\delta = 0.1$. During $J$-evolution we chose the pinning value $a^\dagger$ to be the last value of $a$ produced by the preceding phase of $a$-evolution. Upon reaching a specified value $a^\star$ we carried out an additional $N_{\rm ev}=10^5$ steps of $J$ evolution (with $a^\dagger = a^\star)$, again using $\epsilon=0.05$ and $\delta = 0.1$. \bb{We took $N_{\rm ev}$ large enough that the bound had stopped evolving under $J$-evolution. For the 4-state model the chosen value $10^5$ is much larger than necessary, because the bound stopped evolving after a few hundred trajectories.} We carried out 100 independent simulations, each with a different target value $a^\star$. 

Some of the models produced in this way are shown in \f{fig1}(a), and the associated rate-function bounds are shown as green circles in panel (b). All points $(\tilde{a}_0,J_0)$ on the bound, derived from the typical behavior of the reference models, lie on the exact rate function of the original model, indicating that each reference model's typical dynamics is equivalent to the conditioned rare dynamics of the original model.  Some of the reference models so obtained are shown in panel (a) and in the inset of panel (b). The blue- and gray dashed lines are respectively the Conway-Maxwell-Poisson bound\c{garrahan2017simple} (see Appendix B) and the universal current bound\c{pietzonka2016universal,gingrich2016dissipation}.
\begin{figure}[] 
   \centering
  \includegraphics[width=\linewidth]{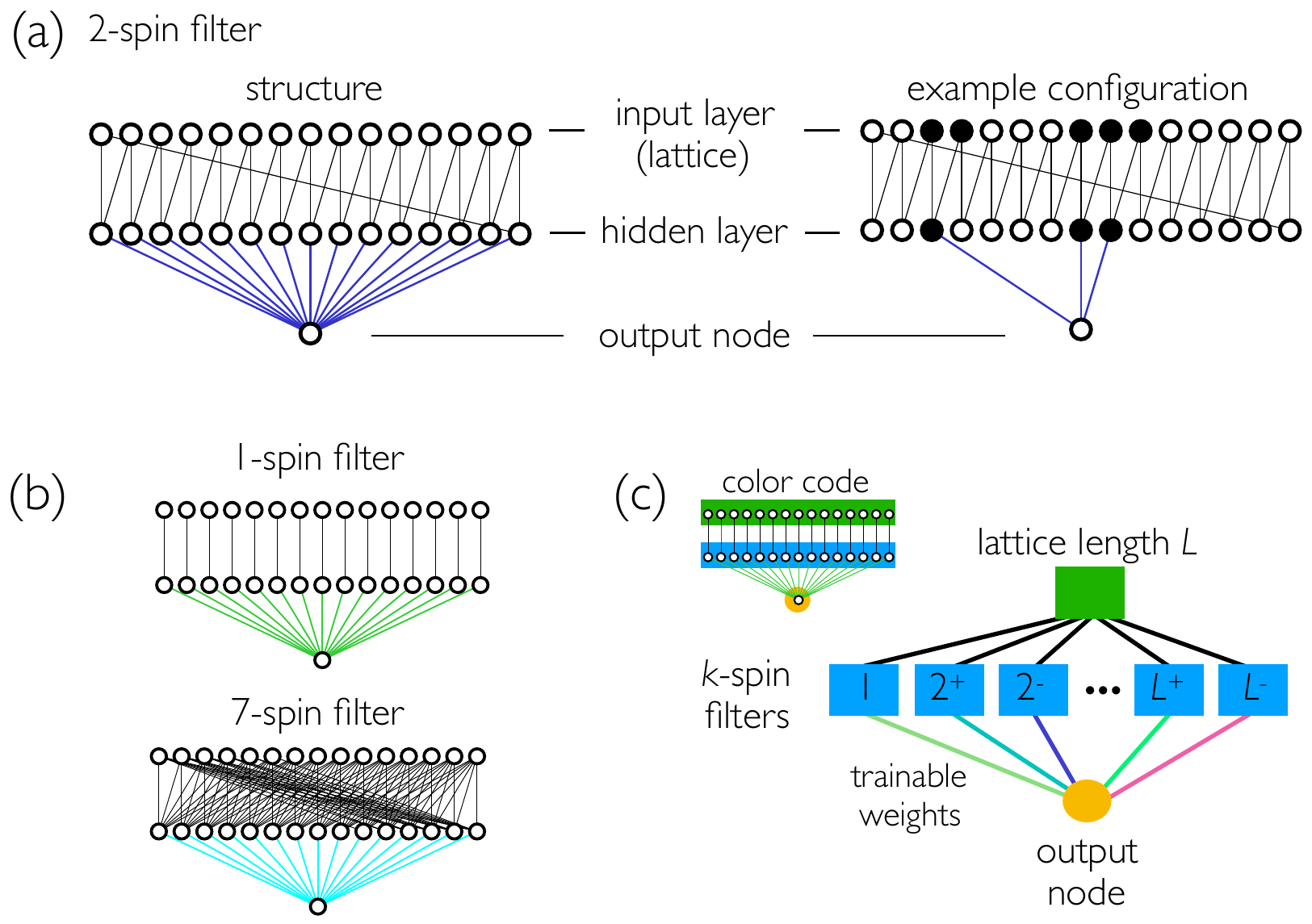} 
   \caption{Sketch of the neural-network reference-model ansatz used to compute the dynamical large deviations of the FA lattice model. (a) The building block of the network is a $k^\alpha$-spin ``filter'', whose hidden nodes activate when the $k$ consecutive spins to which they are attached are all of type $\alpha=\pm $ (periodic boundaries account for the diagonal line between input and hidden layers). Here we show a $2^{+}$-spin filter applied to a lattice of $L=15$ sites; shown right is an example configuration. The output of the filter is the number of hidden nodes that are on (here 3) multiplied by the weights denoted by the blue lines. (b) Structure of a 1-spin filter and a 7-spin filter (lattice size $L=15$). (c) The complete network contains a single hidden layer of $2K$ filters ($K \leq L$), with $2K$ trainable parameters (the colored lines). The network output is the function $f$ displayed in \eq{eff}.}
\label{fig2}
\end{figure}

\section{A neural-network ansatz for models with large state spaces}
\subsection{A lattice model whose state space is small enough to diagonalize}

 In the previous section we saw that evolutionary reinforcement learning using 12 trainable parameters (the 12 rates of the reference model) permits accurate computation of the rate function, i.e. accurate computation of probabilities exponentially small in the trajectory length $T$. However, direct application of rate-based evolution is impractical for models with a large number of rates. To overcome this problem we can encode the rates of the reference model as a neural network, and we illustrate this procedure in this section using the one-dimensional Fredrickson-Andersen (FA) model\c{fredrickson1984kinetic}. 
 \begin{figure*}[] 
   \centering
  \includegraphics[width=\linewidth]{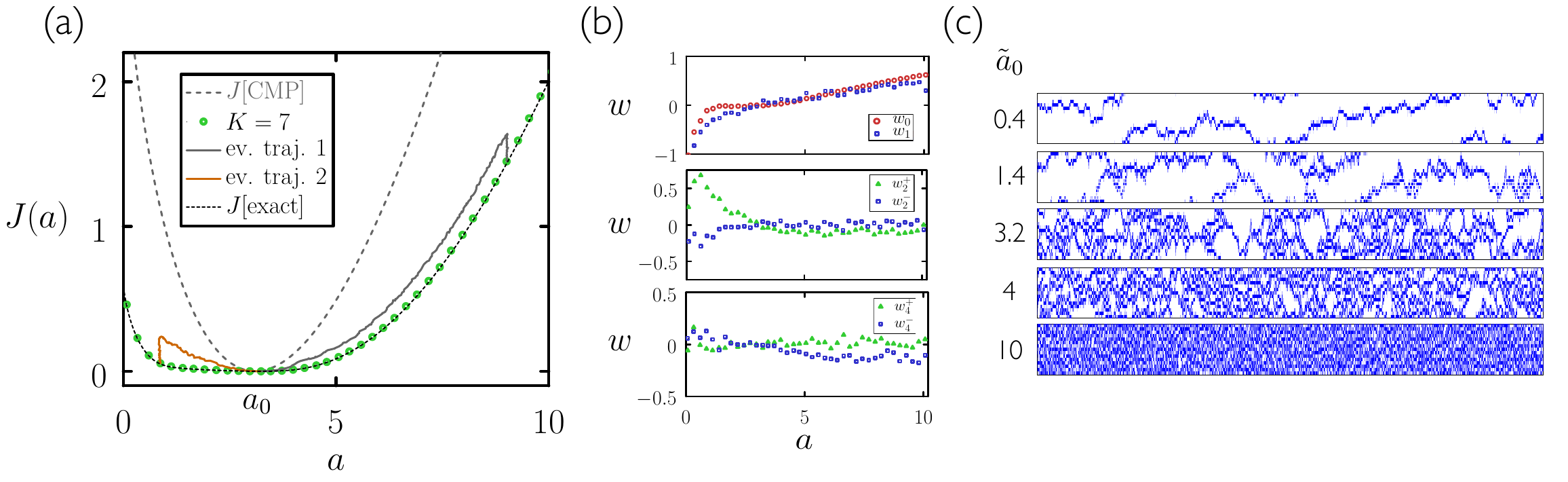} 
   \caption{(a) Evolutionary reinforcement learning using neural-network spin filters up to order $K=7$ (green circles) reproduces the large-deviation rate function $J(a)$ for activity $a$ in the FA model of $L=15$ sites (black). Also shown is the CMP universal activity bound\c{garrahan2017simple} (gray dashed), which results from a set of reference models whose rates are uniform multiples of those of the original model (see Appendix B), and evolutionary trajectories of two neural-network reference models (gray and orange). (b) Values of some of the weights of the neural network \eq{eff} for the reference models that produce the green circles in panel (a). (c) Space (vertical) versus time (horizontal) plots for trajectories of length $T=2 \times 10^3$ for 5 different reference models. Blue pixels indicate up-spins. The typical values of the activity for each model are shown left of the plot; the center reference model is the original model.}
      \label{fig3}
\end{figure*}

The FA model is a lattice model with dynamical rules that give rise to slow relaxation and complex space-time behavior\c{garrahan2002geometrical}. On each site $i$ of a lattice of length $L$ lives a spin $S_i$, which can be up $(+1)$ or down $(-1)$. Up spins (resp. down spins) flip down (resp. up) with rate $1-c$ (resp. $c$) if at least one of their neighboring spins is up; if not, then they cannot flip. We take the dynamical observable $a$ to be the number of configuration changes per unit time, $\alpha_{xy}=1$, often called activity\c{garrahan2007dynamical,garrahan2009first}. To determine the large-deviation rate function $J(a)$ for activity we chose a reference-model parameterization
\beq
\label{rmfa}
\tilde{W}_{xy} = W_{xy} \e^{w_0} \e^{f_y-f_x}.
\eeq 
Here $w_0$ is a parameter that effectively speeds up or slows down the clock\c{jacobson2019direct}~\footnote{In order to calculate the {\em correction} to the bound \eq{j0}, the parameterization using the variable called $\lambda$ in~\cc{jacobson2019direct} is more efficient; to calculate the bound itself the choice ($w_0$ or $\lambda$) makes little difference.}, and $f_x$ is the value in state $x$ of the neural network shown in \f{fig2}. This network is inspired by the convolutional neural networks used to recognize images\c{krizhevsky2012imagenet,lecun1998gradient}, and consists of a set of feature detectors or spin ``filters'' that scan the lattice for specified spin patterns. Here we consider filters called $k^\alpha$, each having $L$ hidden nodes; the output of a hidden node is 1 if the $k$ consecutive spins to which it is attached are all in state $\alpha$, and is zero otherwise (i.e. the activation function is a step function). The network has one hidden layer. The weights connecting the input layer (the lattice) to the hidden layer are unity, and the weights connecting the hidden layer to the output node are denoted $w_k^{\pm}$; these are the trainable parameters of the network. All weights within a filter have the same value, a constraint suggested by the translational invariance of the model. The output of the network is
\beq
\label{eff}
f_x = w_1 g_1({\bm S}_x)+ \sum_{k=2}^K \sum_{\alpha=\pm1} w_k^\alpha g_k^\alpha({\bm S}_x),
\eeq
where ${\bm S}_x$ is the configuration of the lattice in state $x$, and $g_k^\alpha(\cdot)$ returns the number of active hidden nodes in the filter $k^\alpha$ [see \f{fig2}(a)]. The reference model contains $2K$ trainable parameters: $w_0$, $w_1$ (only one type of 1-spin filter is necessary), and $w_2^{\pm},\dots,w_{K}^{\pm}$; note that $K=L$ when all filter types are used. 

The form of \eq{eff} is similar to the multi-parameter auxiliary potential of \cc{nemoto2017finite}, used to improve the convergence of the cloning method\c{giardina2006direct} in order to calculate the large-deviation function of the FA model. The present approach is different, however, in that the calculation is done using direct simulation of a reference model whose parameters are determined by an evolutionary process (rather than using rare-event algorithms such as cloning or transition-path sampling\c{bolhuis2002transition}), and results in the calculation of $J(a)$ directly (rather than its Legendre transform, which in general contains less information\c{touchette2009large}). 

To test the method we considered the FA model with periodic boundary conditions and the parameter choices $c=0.3$ and $L=15$, the latter value being small enough that the exact $J(a)$ can be determined by diagonalization of the model's rate matrix; that function is shown as a black dashed line in \f{fig3}(a). We next introduce the reference model \eq{rmfa}, and do evolutionary reinforcement learning on the weights of the network, as follows.

All neural-network weights $w\in\{w_0, w_1,\{w_K^\alpha\}\}$ of the reference model \eq{rmfa} were initially zero. Each proposed evolutionary move consisted of a shift of each weight by independent Gaussian-distributed random numbers of zero mean and variance $\sigma^2 =10^{-4}$:
\beq
w \to w + \cal{N}(0,\sigma^2).
\eeq
\bb{The parameter $\sigma$ is a learning rate and its effect is similar to other types of learning rate in machine learning, or basic step size in Monte Carlo simulation. We found values of $\sigma$ of order 0.01 to be acceptable.} We ran trajectories for $N=10^5$ events, and recorded the values of \eq{adef} and \eq{j0} after each proposed trajectory. We did $a$-evolution on the parameters $w_0$ and $w_1$ until a specified value $a^\star$ was reached. This procedure was as described for the 4-state model, with the additional restriction that the new bound must be not more than a value $\mu=0.2$ larger than the current bound. That is, the proposed set of weights was accepted if
\beq
\label{acc1}
|\hat{a}-a^\star|<|a-a^\star| \quad {\rm and} \quad \hat{a}<a+\mu.
\eeq
\bb{We introduced the parameter $\mu$ in order to test the effect of replacing the alternating $a$- and $J$-evolution of \s{fourstate} with a ``regularized'' form of $a$-evolution (one that does not allow the bound to grow beyond a particular size in any one step). If $\mu$ was chosen very small (e.g. $ \lesssim 10^{-3}$) then $a$-evolution could not get going at all, because the bound must be allowed to increase in size at some point in the calculation. If $\mu$ was set very large (e.g. of order 10, so that the second requirement in \eq{acc1} was effectively not present), then we observed the effect seen with the gray line in \f{fig_fa_large1}, whereby the bound obtained after $a$-evolution and prior to $J$-evolution was much larger than the exact value of $J$. For intermediate values of $\mu$, such as the value 0.2 chosen here, the bound obtained prior to $J$-evolution was in general close to the exact answer (see the gray and orange lines in \f{fig3}). However, the total CPU time required in the cases of moderate- and large $\mu$ were similar.}

We then did $J$-evolution using a tolerance of $\delta =0.02$ [see \eqq{jev}], for $N_{\rm ev}=3 \times 10^4$ proposed trajectories, with higher-order spin filters applied. We ran 50 simulations, each with a different target value of $a$.

In \f{fig3}(a) we show results of these calculations using spin filters up to order $K=7$. Increasing $K$ from 0 improves the quality of the bound until, for $K \gtrsim 4$, the bound becomes numerically close to the exact answer; see \f{fig4}. \bb{That figure demonstrates that the quality of the bound exceeds that of the few-parameter, physically-motivated ansatz used in \cc{jacobson2019direct}}. The neural network contains many fewer parameters than the model has rates (unlike in many deep-learning studies), and so we do not necessarily expect the bound to be exact. If the bound is good, the exact answer can be calculated by computing a correction term\c{jacobson2019direct}. Here, though, the correction term (the difference between the bound, i.e. the green circles, and the exact answer, i.e. the black dashed line) is very small, indicating that the typical dynamics of this set of reference models is similar to the conditioned rare behavior of the original model. Comparison of these results with the exact result, and with the $(c,\lambda)$-bound from \cc{jacobson2019direct} (\f{fig4}), indicates that rare trajectories of the FA model with parameter $c$ resemble the typical trajectories of versions of the FA model with different values of the parameter $c$, but with slightly different tendencies to display spin domains of different lengths. These tendencies are quantified by the weights of the neural network, some of which are shown in \f{fig3}(b). In panel (c) we show space-time plots of the trajectories of 5 reference models.
\begin{figure}[] 
   \centering
  \includegraphics[width=\linewidth]{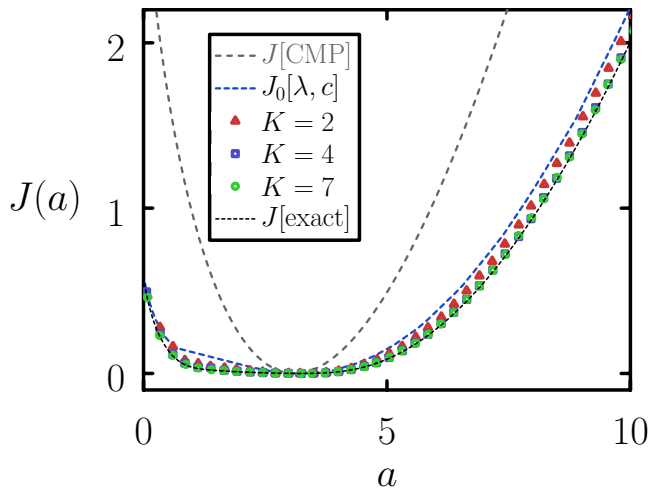} 
   \caption{As \f{fig3}(a), showing results for neural-network spin filters up to order $K=2,4,7$. For $K \gtrsim 4$ the bound is numerically close to the exact answer. Also shown is the CMP universal activity bound\c{garrahan2017simple} (gray), which results from the typical dynamics of a reference model whose rates are uniform multiples of those of the original model, and the $(c,\lambda)$ bound of \cc{jacobson2019direct} (blue). The latter is essentially equivalent to the case $K=1$.}
    \label{fig4}
\end{figure}
\begin{figure*}[] 
   \centering
  \includegraphics[width=0.8\linewidth]{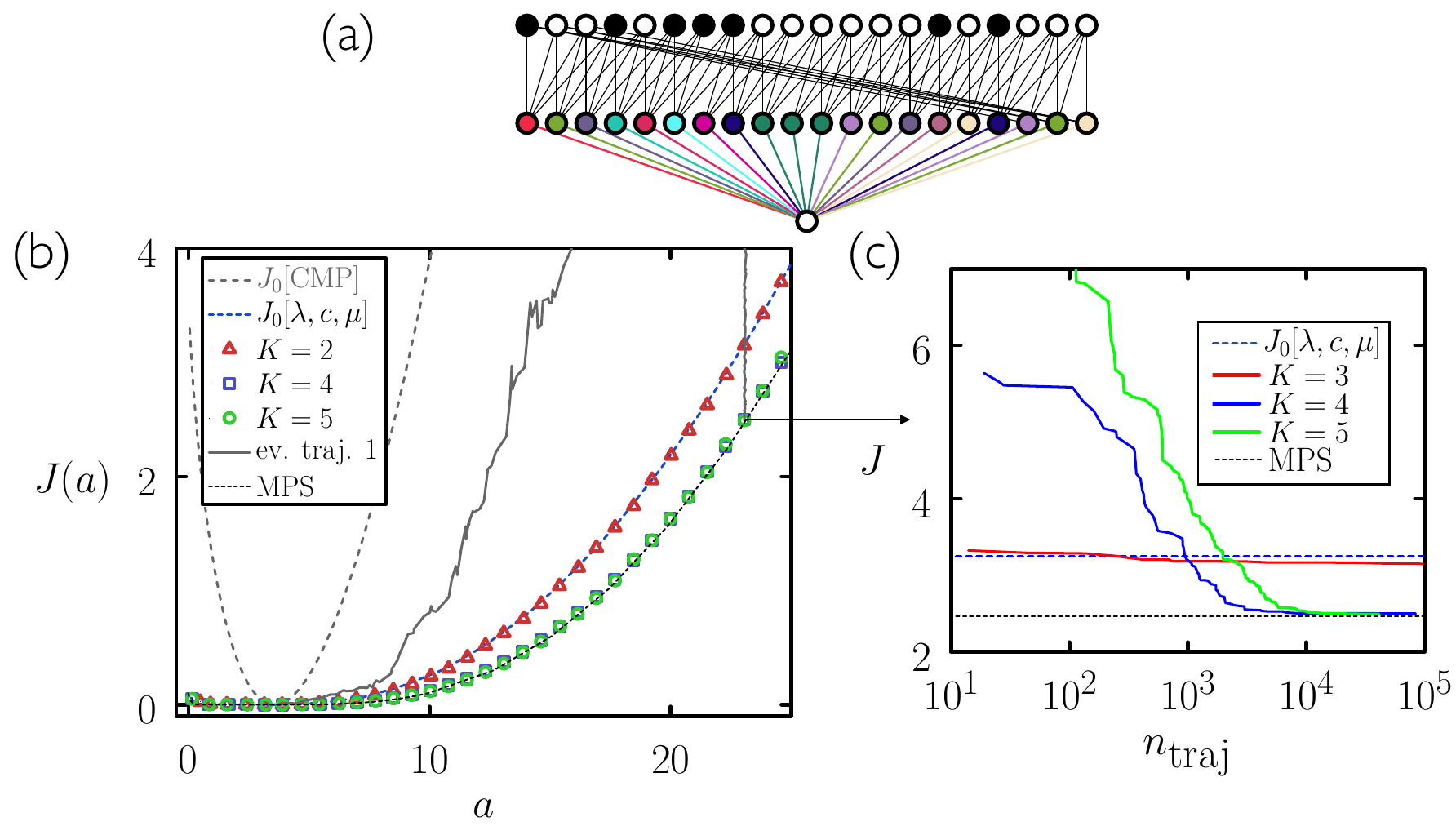} 
   \caption{(a) A spin filter (feature detector) related to those shown in \f{fig2}, but generalized to recognize $2^K$ features in the vicinity of each lattice site (here, for the purpose of illustration, $L=20$ and $K=4$). Each hidden node possesses $2^K$ internal states (identified by colors) and the same number of parameters; the output of the network is \eq{eff2}. (b) Similar to Fig. 3(a), but using the FA model of Ref.\c{banuls2019using} ($L=100,c=0.1$). We show the CMP bound\c{garrahan2017simple} (gray), the three-parameter bound of \cc{jacobson2019direct} (blue), the results of evolutionary learning using the network shown in panel (a), for $K=2,4$, or 5, and the exact answer obtained using matrix product states\c{banuls2019using} (black). Evolutionary learning produces a bound numerically close to the exact answer. We also show one evolutionary trajectory (gray). (c) Bound versus number of trajectories of $J$-evolution for one particular choice of $a$ (labeled by the black arrow connecting panels (b) and (c)), for the cases $K=2,4$ and 5 (the latter corresponds to the gray line in panel (b)).}
      \label{fig_fa_large1}
\end{figure*}

In \f{fig4} we reproduce some of the results shown in \f{fig3}(a), together with results obtained for different values of $K$. \bb{In physical terms the value of $K$ determines the lengthscale over which the dynamical rules of the reference model act. The original model possesses only nearest-neighbor dynamical rules; its dynamics conditioned upon certain values of $a$ involves potentially long-range correlations\c{garrahan2009first}. \f{fig4} shows how closely (in terms of probabilities) the typical dynamics of reference models whose dynamical rules possess $K$-spin correlations approximate this conditioned dynamics: $K \approx 4$ is sufficient to closely approximate the rate function.}

\subsection{A lattice model whose state space is too large to diagonalize}

With proof of principle demonstrated using models whose state space is small enough to solve by matrix diagonalization, we show in \f{fig_fa_large1} that bounds produced by evolutionary learning can closely approximate rate functions whose calculation requires state-of-the-art numerical methods. For this purpose we chose the 100-site FA model of Ref.\c{banuls2019using} (for $c=0.1$), whose state space is large enough that state-of-the-art methods are needed to compute its large-deviation rate function. This FA model has open boundary conditions.

Initial tests with this larger FA model, done using evolutionary learning on the network shown in \f{fig2} (with $K=5$), produced a bound that was visibly less close to the exact answer than in the case $L=15$, suggesting the need for a neural-network ansatz able to detect more detailed features. We therefore replaced the network shown in \f{fig2} with the one shown in \f{fig_fa_large1}(a). This new network is capable of learning which features (spin patterns) are most significant; by contrast, the network shown in \f{fig2} searches only for homogenous blocks of spins.

Each hidden node $i=1,2,\dots L$ in the new network couples to $K$ input nodes (lattice sites), and takes one of $2^K$ values. This value, called $h_x(i)$ in microstate $x$, is determined by the state of the $K$ spins to which is it connected, via
\beq
h_x(i) = \sum_{m=0}^{K-1}2^m\left( \frac{1+S^x_{i+m}}{2}\right).
\eeq
The output of the network in microstate $x$ is then 
\beq
\label{eff2}
f_x=\sum_{i=1}^L w_{h_x(i)},
\eeq
where the $2^K$ weights $w_{h_x(i)}$ are, along with $w_0$, the trainable parameters of the model.  The reference-model ansatz is again \eq{rmfa}, but now with \eq{eff2} replacing \eq{eff}. 

We ran 40 evolutionary simulations, each with a different target value of $a$ between 0.1 and 30 (the typical $a$ of the original model is approximately 3.5). We turned on the neural network from the start, and used trajectories of $N = 2 \times 10^5$ events. We did $a$-evolution (using \eqq{simple})  to generate the desired values of $a$, and then did $J$-evolution for $ \approx 3 \times 10^4$ proposed trajectories. Results are shown in \f{fig_fa_large1}(b): for $K \gtrsim 4$, the bound produced is inexact, but numerically close to the exact answer. In panel (c) we show the evolution of the bound as a function of the number of evolutionary steps $n_{\rm traj}$. As a guide to CPU consumption, 100 trajectories of $N = 2 \times 10^5$ events (each followed by a neural-network mutation step) take 5, 17, and 31 seconds for the cases $K=2,4$ and 5, respectively, on a 3.1 GHz Intel Core i7 processor (and so the total simulation time for the case $K=5$ was of order 4 hours on that processor). 

\bb{We note that we used slightly different variants of the $a$- and $J$-evolution protocols for each of the 4-state model and the small- and large FA models (each protocol is detailed above), in order to explore the effect of changing protocol. We did not find one protocol to be obviously better than the others, suggesting that a number of different evolutionary strategies can be used to tackle these problems.}  

\section{Conclusions}  In previous work we showed how to calculate dynamical large-deviation rate functions using a variational ansatz for rare dynamics (VARD) \c{jacobson2019direct}. The first step of the VARD method is to calculate a rate-function bound, derived from the typical behavior of the ansatz, and we showed in that paper that ans\"atze containing a few parameters motivated by physical insight produced tight bounds for a set of models taken from the literature. In this paper we have shown that multiparameter ans\"atze, in the form of a relatively simple neural network (``VARDnet''), combined with evolutionary reinforcement learning, produce even tighter bounds on dynamical large-deviation rate functions for three such models. In these cases no physical insight into the model under study was required. The second step in the VARD method is to calculate a correction term in order to turn the bound into the exact rate function; here, for the three models considered, the discrepancy between bound and exact answer (obtained by other means) is so small that for the purposes of plotting the rate function no correction is required. In the case of \f{fig_fa_large1}, calculation of the large-deviation rate function for the model in question requires state-of-the-art methods\c{banuls2019using}. 

In treating the two lattice models we have introduced neural networks as reference-model ans\"atze for the rare behavior of each; the question of which network is best for a particular model and application is an open one. We used the single-layer architectures shown in \f{fig2} and \f{fig_fa_large1}, partly because the rates for reference-model spin flips then depend only on the states of the feature detectors to which a spin is attached (here a number of order 5), and this allows relatively efficient and rapid updating of rate tables during the course of a continuous-time Monte Carlo simulation. A natural next step would be to apply a deeper network during later stages of evolution (e.g. once the evolutionary trajectories in \f{fig_fa_large1}(c) have reached their plateaux). Doing so would make for more costly simulation, but would allow each reference-model rate to be informed by the state of the entire lattice, thereby increasing the descriptive power of the ansatz.

The approach described here does not rely on the formal results of large-deviation theory, making it complementary to the growing body of methods based on such results\c{giardina2006direct,touchette2009large,garrahan2009first,chetrite2015variational,jack2015effective,ray2018exact,nemoto2017finite,ferre2018adaptive}. More generally, the present approach can be adapted to treat other physical problems that involve time- or path-extensive quantities; one example is molecular self-assembly, whose outcome depends in some potentially complex way on the entire history of the interactions of a set of molecules\c{whitelam2019learning}.

\section{Acknowledgments} 
 We thank Hugo Touchette for comments. This work was performed as part of a user project at the Molecular Foundry, Lawrence Berkeley National Laboratory, supported by the Office of Science, Office of Basic Energy Sciences, of the U.S. Department of Energy under Contract No. DE-AC02--05CH11231. D.J. acknowledges support from the Department of Energy Computational Science Graduate Fellowship. I.T. performed work at the National Research Council of Canada under the auspices of the AI4D Program.

\appendix
\section{Large deviations by change of model}
\label{com}
For completeness we present the derivation of \eqq{j0} of the main text, which follows straightforwardly from the definition of the probability distribution. The derivation follows \cc{jacobson2019direct} with minor notational changes. For more on the ideas of dynamic importance sampling see e.g. Refs.\c{glynn1989importance,sadowsky1990large,bucklew1990monte,bucklew1990large,asmussen2007stochastic,juneja2006rare,bucklew2013introduction,touchette2009large} and \cc{chetrite2015variational} (esp. Section 5).

Consider a continuous-time dynamics on a set of discrete states, defined by the master equation\c{binder1986introduction}
\beq
\label{me}
\partial_t P_x(t) = \sum_{y\neq x} W_{yx} P_y(t)- R_x P_x(t).
\eeq
Here $P_x(t)$ is the probability that the system is in (micro)state $x$ at time $t$, $W_{xy}$ is the rate for passing from state $x$ to state $y$, and $R_x = \sum_{y \neq x} W_{xy}$ is the escape rate from $x$. A standard way of simulating \eq{me} is as follows\c{gillespie1977exact}: from state $x$, choose a new state $y$ with probability
\beq
\label{one}
p_{xy} = \frac{W_{xy}}{R_x},
\eeq
and a time increment $\Delta t$ from the distribution
\beq
\label{two}
p_{x}(\Delta t) = R_x \e^{-R_x \Delta t}.
\eeq
The dynamics defined by \eq{one} and \eq{two} generates a trajectory $\omega=x_0 \to x_1 \to \dots \to x_{N(\omega)}$ consisting of $N(\omega)$ jumps $x_n \to x_{n+1}$ and associated jump times $\Delta t_n$. Associated with an ensemble of trajectories of length $T$ is the probability distribution
\beq
\label{av}
\rho_T(A)=\sum_{\omega} p(\omega) \delta{(T)} \delta{(A)}
\eeq
of a time-extensive dynamical observable 
\beq
\label{ay}
A(\omega) = \sum_{n=0}^{N(\omega)-1} \alpha_{x_n x_{n+1}}.
\eeq
In these expressions $\delta(X) \equiv \delta{(X(\omega) - X)}$ specifies a constraint on the trajectory, $\alpha_{xy}$ is the change of $A$ upon moving from $x$ to $y$, and $A(\omega)$ is the sum of these quantities over a single trajectory $\omega$. We define $a(\omega) \equiv A(\omega) / T(\omega)$ as the time-intensive version of $A$. $T(\omega)$ is the elapsed time of trajectory $\omega$, and $p(\omega)$ is the probability of a trajectory $\omega$, proportional to a product of factors \eq{one} and \eq{two} for all jumps of the trajectory.

Fluctuations of $a$ are quantified by $\rho_T(A)$, which for large $T$ often adopts a large-deviation form\c{den2008large,touchette2009large}
\beq
\label{ld}
\rho_T(A) \approx \e^{-T J(a)}.
\eeq 
Direct evaluation of \eq{av} using the dynamics \eq{one} and \eq{two} leads to good sampling of $J(a)$ near the typical value $a_0$, where $J(a_0)=0$, and poor sampling elsewhere. To overcome this problem we can introduce a reference dynamics
\beq
\label{three}
\tilde{p}_{xy} = \frac{\tilde{W}_{xy}}{\tilde{R}_x},
\eeq
and
\beq
\label{four}
\tilde{p}_{x}(\Delta t) = \tilde{R}_x \e^{-\tilde{R}_x \Delta t},
\eeq
in which $\tilde{W}_{xy}$ is a modified version of the rate of the original model, and $\tilde{R}_x \equiv \sum_y \tilde{W}_{xy}$. Let $\tilde{p}(\omega)$ be the trajectory weight of the reference dynamics, proportional to a product of factors \eq{three} and \eq{four} for all jumps of the trajectory. We write 
\beq
\av{\cdot}^a \equiv \sum_\omega p(\omega) (\cdot) \delta{(T)} \delta{( aT)}
\eeq
and
\beq
\av{\cdot}_{\rm ref}^a \equiv \sum_\omega \tp(\omega) (\cdot) \delta{(T)} \delta{(a T)}
\eeq
for the ensemble averages over trajectories (having length $T$ and observable $A=aT$) of the original and reference models, respectively. We can then write \eq{av} as
\bea
\label{av2}
\rho_T(A) &=&\av{1}^a \nonumber \\ &=& \av{\e^{T q(\omega)}}_{\rm ref}^a\\
\label{av22}
&=&\e^{T \av{q(\omega)}_{\rm ref}^a} \av{\e^{T \delta q(\omega)}}_{\rm ref}^a.
\eea
Here $\e^{T q(\omega)}=p(\omega)/\tilde{p}(\omega) $ is the reweighting factor (also known as the likelihood ratio or Radon-Nikodym derivative\c{chetrite2015variational,bucklew2013introduction}). We have
 \beq
 \label{phi-def}
 q(\omega) = T^{-1} \ln \frac{p(\omega)}{\tilde{p}(\omega) } =-T^{-1} \sum_{n=0}^{N-1} q_{x_n x_{n+1}},
 \eeq
 where
\beq
\label{phi_micro}
q_{x_n x_{n+1}} = \ln \frac{W_{x_n x_{n+1}}}{\tilde{W}_{x_n x_{n+1}}}-\tilde{\Delta t}_n (R_{x_n} - \tilde{R}_{x_n}).
\eeq
Here $\tilde{\Delta t}_n = -\ln \eta/\tilde{R}_{x_n}$ is the jump time of the reference model ($\eta$ is a random number uniformly distributed on $(0,1]$). In \eq{av22} the quantity $\delta q(\omega) \equiv q(\omega)-\av{q(\omega)}_{\rm ref}^a$.

Taking logarithms of \eq{av22} and the large-$T$ limit gives us
\beq
\label{final}
J(\ta0) = J_0(\ta0)+J_1(\ta0),
\eeq
where
\beq
\label{bound}
J_0(\ta0)= -\av{q(\omega)}_{\rm ref}^{\ta0}
\eeq
and 
\beq
\label{correction}
J_1(\ta0) =-\frac{1}{T} \ln \av{\e^{T \delta q(\omega)}}_{\rm ref}^{\ta0}.
\eeq
In these expressions $\ta0$ is the typical value of $a$ for the reference model. The term \eq{bound} is by Jensen's inequality an upper bound on the piece of the rate function $J(a)$ at the point $a=\ta0$, i.e
\beq
\label{jay}
J(\ta0) \leq J_0(\ta0).
\eeq
The bound can be determined by computing the values of \eq{ay} and \eq{phi-def} for a suitably long reference-model trajectory. If the reference model's typical dynamics is similar to the conditioned rare dynamics of the original model (something we generally do not know in advance), then the bound $J_0(\ta0)$ will be tight, and if it is tight enough the exact value of $J(\ta0)$ can be calculated by sampling the (slightly) atypical behavior of the reference model\c{jacobson2019direct}. In the main text we show that evolutionary reinforcement learning can generate reference models for which the correction term is very small. The optimal reference model, called the driven or auxiliary process\c{chetrite2015variational,jack2015effective,bucklew2013introduction,touchette2009large}, is one for which the bound is exact, meaning that its typical behavior is equivalent to the conditioned rare behavior of the original model. 

\section{The CMP universal activity bound}
\label{sec_cmp}

\begin{figure}[] 
   \centering
  \includegraphics[width=\linewidth]{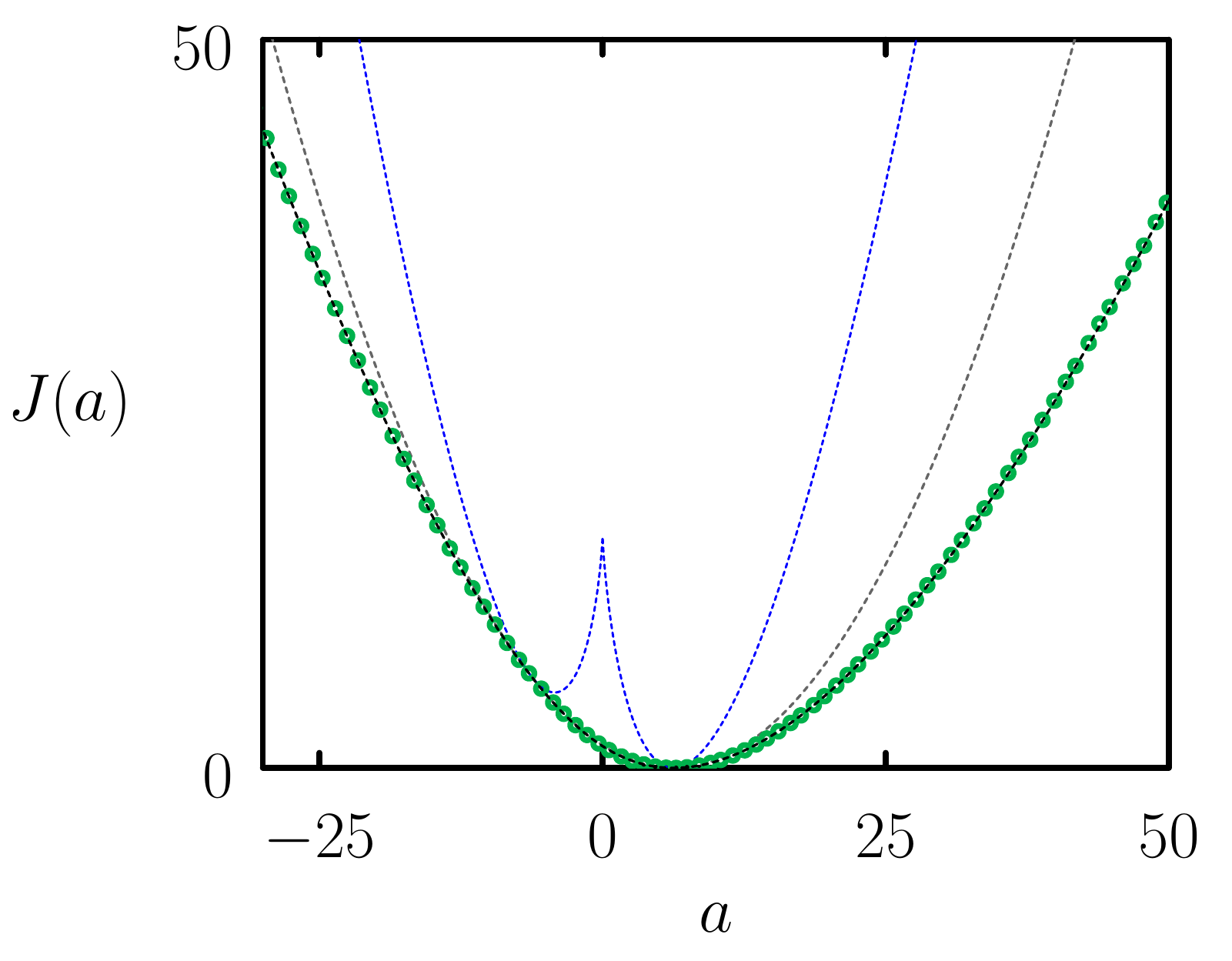} 
   \caption{Supplement to Fig. 1 of the main text, the large-deviation rate function $J(a)$ for entropy production $a$ in a 4-state model (black and green). We show as a blue dashed line the CMP universal activity bound\c{garrahan2017simple} and its time reverse, \eqq{j03}, which together provide a rudimentary bound on any current. Shown in gray is the universal current bound\c{pietzonka2016universal,gingrich2016dissipation}.}
   \label{fig_bounds}
\end{figure}

The Conway-Maxwell-Poisson (CMP) formula
\beq
\label{cmp}
J_{\rm CMP}(a) = \frac{k_0}{a_0} \left( a \ln \frac{a}{a_0}+ a_0 -a\right),
\eeq
gives a bound on the large-deviation rate function $J(a)$ for any non-decreasing counting observable $a$; here $a_0$ is the typical value of the observable, and $k_0$ is the typical dynamical activity $k$ (the total number of configuration changes per unit time). \eqq{cmp} was derived in \cc{garrahan2017simple} from Level 2.5 of large deviations\c{maes2008canonical,bertini2015large}, and we have used this form in \f{fig3}, \f{fig4}, and \f{fig_fa_large1} (for the case $a=k$).

We note here that the CMP formula can be straightforwardly derived from the generic bound \eq{j0}, without using the result known as Level 2.5 of large deviations.  Let $a_0$ and $k_0$ be the typical activities produced by an original model $W_{xy}$. Then a reference model $\tilde{W}_{xy} = \gamma W_{xy}$, whose rates are uniformly rescaled versions of those of the original model, will produce typical activities $\gamma a_0$ and $\gamma k_0$ (a uniform rescaling of rates does not affect the choice of new state, i.e. $\tilde{W}_{xy}/\tilde{R}_x=W_{xy}/R_x$, and so the reference model will visit the same set of states as the original model, just faster or slower).  The reference-model escape rate is then $\tilde{R}_x = \gamma R_x$. In \eq{j0} we assume the long-time, steady-state limit, and so replace the fluctuating jump time $\tilde{\Delta t}_n$ with its mean $1/\tilde{R}_{x_n}$,  giving
\bea
\label{j02}
J_0(\tilde{a}_0)&=& T^{-1} \sum_{n=0}^{N-1} \left(\ln \gamma +\frac{1-\gamma}{\gamma} \right )\nonumber\\
&=&\tilde{k}_0\left(\ln\frac{\tilde{a}_0}{a_0} +\frac{a_0-\tilde{a}_0}{\tilde{a}_0} \right ) \nonumber \\
&=&\frac{\tilde{k}_0}{\tilde{a}_0}\left(\tilde{a}_0\ln\frac{\tilde{a}_0}{a_0} +a_0-\tilde{a}_0 \right ) \nonumber \\
&=&\frac{k_0}{a_0}\left(\tilde{a}_0\ln\frac{\tilde{a}_0}{a_0} +a_0-\tilde{a}_0 \right ),
\eea
where $\tilde{a}_0= \gamma a_0$ and $\tilde{k}_0= \gamma k_0$ are respectively the mean value of $a$ and $k$ for the reference model ($a_0$ and $k_0$ are the analogous quantities for the original model). \eqq{j02} is the bound associated with the single reference model whose rates are $\tilde{W}_{xy} = \gamma W_{xy}$. By choosing different values of $\gamma$ we create a family of reference models, each with a distinct typical behavior, and so we can replace $\tilde{a}_0$ in \eqq{j02} with the general $a$; doing so, we recover \eqq{cmp}, the CMP bound. 

The derivation leading to \eq{j02} specifies only that $a=A/T$ be derived from a time-extensive quantity $A$, so that rescaling all rates by a factor $\gamma>0$ changes the typical value of the observable, $a_0$, to $\gamma a_0$. Thus the CMP bound applies to {\em any} time-extensive quantity, including currents, not just non-decreasing counting variables. This fact justifies its inclusion in \f{fig1}, where we consider entropy production (a current). However, the CMP bound does not address the $a<0$ sector, which cannot be accessed if the typical value of the observable of the original model is $a_0>0$ (because any reference model obtained under a rescaling of rates has $\tilde{a}_0=\gamma a_0 >0$). 

A simple way to produce a bound pertaining to the $a<0$ sector is to use the $\gamma$-rescaling on the {\em time-reversed} version of the original model. To see this, we proceed as follows. Consider the reference model obtained by rescaling the rates of the original model, $W_{xy}$, by the exponential of (minus) the entropy production:
\bea
\label{oneb} \tilde{W}_{xy}&=&\e^{-\sigma_{xy}} W_{xy} \\
\label{twob} &=& \frac{\pi_y}{\pi_x} \frac{p_{yx}}{p_{xy}} W_{xy} \\
\label{threeb} &=&\frac{R_x}{\pi_x} \pi_y p_{yx}.
\eea
Here we are using standard notation for Markov chains: $p_{xy}=W_{xy}/R_x$ is the probability of moving to (micro)state $y$, given that we are in state $x$; $R_x = \sum_y W_{xy}$ is the escape rate from $x$; and $\pi_x$ is the invariant measure, which satisfies
\beq
\label{im}
\pi_x= \sum_y \pi_y p_{yx}.
\eeq
Summing \eq{threeb} over $y$ and using \eq{im} shows that the escape rate of the reference model is equal to that of the original:
\beq
{\tilde R}_x = \sum_y W_{xy} = \sum_y {\rm \eq{threeb}}= \frac{R_x}{\pi_x} \pi_x = R_x.
\eeq
Then upon dividing \eq{twob} by ${\tilde R}_x = R_x$ we have 
\beq
{\tilde p}_{xy} =  \frac{\pi_y}{\pi_x} p_{yx},
\eeq
and so this reference model generates the time-reversed Markov chain\c{hastings1970monte}. If the observable $a$ is a current, odd under time reversal, then the typical value of the observable in the reference model is $\tilde{a}_0 = -a_0$. 

To determine the value of the bound associated with the time-reversed model we inset \eq{oneb} into \eq{j0}, giving
\beq
\label{j0rev}
J_0 =-T^{-1} \sum_{n=0}^{N-1} \sigma_{x_n x_{n+1}} = -\tilde{\sigma}_0 = \sigma_0. 
\eeq
Thus choosing the time-reversed model to be the reference model gives as a bound a single point $(-a_0,\sigma_0)$ on the rate function of any current $a$; here $a_0$ and $\sigma_0$ are the typical values of the current and the entropy production rate in the original model.

If we now apply a $\gamma$-rescaling to the time-reversed model we create a family of reference models with rates
\beq
{\tilde W}_{xy}  = \gamma \e^{-\sigma_{xy}} W_{xy}.
\eeq 
Using \eq{j0} and the results \eq{j02} and \eq{j0rev} it is straightforward to show that the bound associated with this family of reference models is 
\beq
\label{j03}
J_0(a) = \frac{\sigma_0}{a_0} |a| + \frac{k_0}{a_0} \left(|a|\ln\frac{|a|}{a_0} +a_0-|a|\right ).
\eeq
Hence one bound on any current $a$ is provided by the combination of \eq{j02} (with $\tilde{a}_0 \to a$) for $a\geq 0$ and \eq{j03} for $a<0$. We show this bound (for the choice $a=\sigma$ for the 4-state model) as a blue dotted line in \f{fig_bounds}. The double-well form results from the fact that the associated family of reference models is a glued-together combination of forward and time-reversed `original' models with uniformly rescaled rates. Comparison with the universal current bound\c{pietzonka2016universal,gingrich2016dissipation} (gray dotted line) shows the latter to derive from a different family of models (see Figs. 2 and 3 of \cc{jacobson2019direct} for a comparison between the universal current bound and the bounds produced by other families of reference models).
%

\end{document}